\documentclass[conference]{IEEEtran}
\IEEEoverridecommandlockouts
\usepackage{cite}
\usepackage{amsmath,amssymb,amsfonts}
\usepackage{algorithmic}
\usepackage{graphicx}
\usepackage{textcomp}
\usepackage{xcolor}
\usepackage{float}
\usepackage{multirow}
\usepackage{fancyhdr}

\fancypagestyle{firstpage}{
    \fancyhf{} 
    \fancyhead[L]{2024 27th International Conference on Computer and Information Technology (ICCIT)\\
    20-22 December 2024, Cox’s Bazar, Bangladesh}
    \fancyfoot[L]{979-8-3315-1909-4/24/\$31.00 \, \copyright\, 2024 IEEE}
}

\title{ EmoFormer: A Text-Independent Speech Emotion Recognition using Hybrid Transformer-CNN model \\

\author{
    \IEEEauthorblockN{Rashedul Hasan, Meher Nigar, Nursadul Mamun, Sayan Paul}
    \IEEEauthorblockA{
       Robust Speech Processing Laboratory (RSPL)\\ Chittagong University of Engineering and Technology, Chittagong
       \\ (rashedulhasan2079, mnigar6653)@gmail.com, nursad.mamun@cuet.ac.bd, sayanpaul1952@gmail.com
       }
}

}

\begin{document}

\maketitle
\thispagestyle{firstpage}

\begin{abstract}
Speech Emotion Recognition is a crucial area of research in human-computer interaction. While significant work has been done in this field, many state-of-the-art networks struggle to accurately recognize emotions in speech when the data is both speech and speaker-independent. To address this limitation, this study proposes, EmoFormer, a hybrid model combining CNNs (CNNs) with Transformer encoders to capture emotion patterns in speech data for such independent datasets. The EmoFormer network was trained and tested using the Expressive Anechoic Recordings of Speech (EARS) dataset, recently released by META. We experimented with two feature extraction techniques—MFCCs and x-vectors. The model was evaluated on different emotion sets comprising 5, 7, 10, and 23 distinct categories. The results demonstrate that the model achieved its best performance with five emotions, attaining an accuracy of 90\%, a precision of 0.92, a recall, and an F1-score of 0.91. However, performance decreased as the number of emotions increased, with an accuracy of 83\% for seven emotions compared to 70\% for the baseline network. This study highlights the effectiveness of combining CNNs and Transformer-based architectures for emotion recognition from speech, particularly when using MFCC features.

\end{abstract}


\section{Introduction}
Speech Emotion Recognition (SER) has become a critical area of research in human-computer interaction \cite{ramakrishnan2013speech}. The applications of SER range from healthcare to customer service and virtual assistants. Emotions play a fundamental role in the communication sector and the ability to automatically detect them from speech enhances the effectiveness of interactions between humans and machines.\cite{avro2024Emotech} However, SER presents unique challenges due to the variability in vocal tone, pitch, speed, and context. This can significantly affect the emotional content of speech \cite{wani2021comprehensive}.

Over the years, various methods have been employed for SER. Early research relied on traditional machine learning approaches like Support Vector Machines (SVMs) and Hidden Markov Models (HMMs). In \cite{koduru2020feature}, the proposed system enhances SER by extracting features using discrete wavelet transform (DWT), pitch, energy, and zero crossing rate. Here, the decision tree classifier outperforms other classifiers like SVM and linear discriminant analysis (LDA). In \cite{mao2019revisiting}, the study compared three HMM-based architectures for SER where SGMM-HMMs performed best on multiple datasets. However, deep learning has transformed the field, allowing for more automated and sophisticated feature extraction and classification techniques. In \cite{kerkeni2019automatic}, a comparative study was carried out utilizing MFCC and modulation in spectral features for SER systems. In this study, RNN performed better than multiple linear regression (MLR) and SVM. Despite this, the RNN approach struggled with computational efficiency, particularly for large datasets. The work in \cite{mirsamadi2017automatic} presents a deep recurrent neural network with a novel attention-based feature pooling strategy for improved automatic emotion recognition from speech. In \cite{zhao2019speech}, the study developed 1D and 2D convolutional neural networks (CNN) with long short-term memory (CNN-LSTM) networks to learn local and global emotion features from speech. The model outperformed traditional methods like deep belief network (DBN) and CNN. The authors in \cite{anvarjon2020deep, sarker2023text} proposed a lightweight CNN-based SER model with low computational complexity and high accuracy. More recently, transformer-based architectures have emerged, providing a powerful means to capture contextual relationships across entire speech sequences. In \cite{he2023multiple}, the authors proposed a novel SER system using a cross-attention transformer to fuse raw waveform data, spectrogram, and MFCC features. However, the reliance on multiple input modalities increased the model’s complexity and make it challenging to apply in real-time applications. The paper \cite{morais2022speech} introduced a modular End-to-End SER system leveraging self-supervised features and demonstrated its ability to achieve SOTA results using only the speech modality. However, it is limited by the lack of multitask learning and multimodal integration. The paper in \cite{aftab2022light} proposes a lightweight FCNN for speech emotion recognition, designed to be efficient for systems with limited hardware resources. Although it achieved competitive performance compared to state-of-the-art models, the model's smaller size makes it more suitable for embedded applications.

Despite advancements in SER, most studies remain focused on either traditional machine learning methods or deep learning models that utilize single feature sets. While these approaches show effectiveness, they often struggle with speech and speaker variability and fail to fully capture the temporal and contextual dynamics of emotional speech.  Variability in speech arises from differences in accents, speaking styles, and environmental factors. These limitations highlight the need for a more comprehensive approach that integrates diverse feature representations and advanced deep learning architectures.

To address these challenges, this study proposes a hybrid approach that leverages both MFCC and x-vector features, combined with CNN and transformer architectures.MFCC features effectively capture the spectral properties of speech and x-vector features provide robust speaker-independent representations. The integration of CNNs and transformer-based encoders enables the system to extract spatial features and model long-range temporal and contextual dependencies and addresses both speech variability and the dynamic nature of emotions. 

The contributions of this work are twofold: first, the effective use of MFCC features to capture the spectral properties of speech. Second, the introduction of transformer-based encoders to model long-range contextual relationships and speaker variability in speaker-independent datasets. By fusing these diverse feature sets with advanced deep learning architectures, our system aims to significantly enhance the accuracy of emotion recognition in speech.


The organization of this paper is as follows: Section II offers a comprehensive overview of the data preprocessing and feature extraction methods employed. Section III details the development of the hybrid models implemented in this study. Part IV presents the results and analyzes the outcomes of emotion recognition. Lastly, Section V wraps up the paper by addressing the limitations encountered and suggesting possible directions for future research. 

\section{Methodology}

\subsection{Dataset}

\begin{figure}[t]
  \centering
  \includegraphics[width=0.5\textwidth]{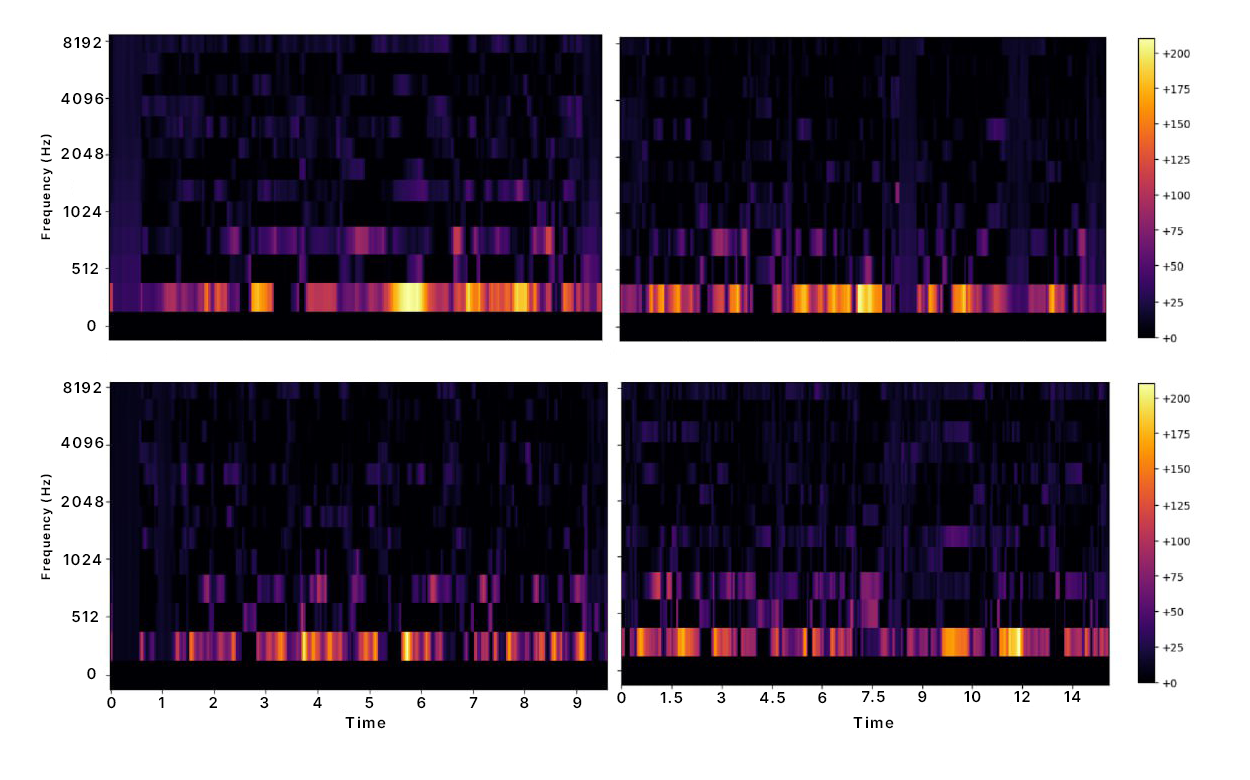}
  \caption{MFCC features for different emotions such as Adortion, Sadness, Anger, and Neutral.}
  \label{fig: mfcc feature}
   \vspace{-20pt}
\end{figure}

In speech emotion recognition, state-of-the-art models often face challenges with text-independent datasets due to variability in spoken content. To tackle this issue, our study introduces a hybrid model using the Expressive Anechoic Recordings of Speech (EARS) dataset, designed for both subject- and text-independent emotion recognition.

The recently released EARS dataset, provided by META, includes 749 audio files from 107 speakers, amounting to 100 hours of clean, anechoic speech. It covers 23 distinct emotions, with 107 samples per emotion, all recorded at a sampling rate of 48 kHz. The dataset features high speaker diversity, spanning various ethnic backgrounds and age groups ranging from 18 to 75 years. Audio durations within the dataset vary significantly, with the shortest file lasting 1.92 seconds and the longest extending to 30.59 seconds, while the average file length is 14.51 seconds. This wide range of durations not only introduces natural variability but also enhances the diversity of speech samples, improving model robustness. Though the original dataset included 23 distinct emotions, we experimented with different sets of emotions in the context of this work. This even distribution helps prevent bias toward any particular emotion during the development of the recognition models.

\subsection{Data Augmentation}

Due to the limited dataset size, data augmentation techniques were applied to the training set to enhance its diversity. To ensure consistency during augmentation, all audio files were resampled at 16,000 Hz. The speech signals were then time-stretched by factors of 0.9 and 1.1, simulating variations in speech delivery rate without altering the pitch. Additionally, the pitch was adjusted by ±2 semitones, introducing vocal pitch variations that naturally occur due to different user characteristics or emotional states. Lastly, all audio files were either padded or truncated to a fixed length of 15 seconds for uniformity.

\subsection{Feature Extraction}

\subsubsection{Mel-frequency cepstral coefficients}
Mel-frequency cepstral coefficients (MFCC) are features used in speech processing to capture the most important aspects of the speech signal. MFCCs are computed by transforming a raw audio signal into a more compact representation based on how humans perceive sound. 

At first, the signal $x[n]$ is pre-emphasized through a filter to boost high frequencies:
\begin{equation}
y[n] = x[n] - \alpha x[n-1]
\end{equation}
where $0.9 < \alpha < 1$. 

The signal is then divided into overlapping frames, and each frame is multiplied by a window function. The Fourier Transform converts each frame from the time domain to the frequency domain, giving the magnitude spectrum:
\begin{equation}
X_k[m] = \sum_{n=0}^{N-1} x_k[m] e^{-j 2\pi mn / N}
\end{equation}
where $0 < m < M$. 

Then, the spectrum is passed through filters spaced on the Mel scale:
\begin{equation}
\text{Mel}(f) = 2595 \log_{10}\left(1 + \frac{f}{700}\right)
\end{equation}

The logarithm of the filter bank output is computed to mimic the ear’s sensitivity to loudness. Finally, the Discrete Cosine Transform (DCT) is applied to the logarithm of the Mel-filtered energies to decorrelate them and obtain the MFCCs.
Here, the MFCC features for various emotions extracted in this work are illustrated in Fig.~\ref{fig: mfcc feature}. The features were derived from audio using 13 coefficients, which represent the spectral characteristics of the speech signal. These coefficients capture speech features over time, with each time frame associated with a set of MFCC values.

To enhance temporal pattern recognition, the extracted MFCC features were divided into overlapping segments, each consisting of 469-time frames with an overlap of 128 frames. This segmentation enables the model to effectively capture the temporal patterns critical for emotion recognition.

\subsubsection{X-vector Feature Extraction}\label{AA}

X-vector extraction is a deep learning-based method commonly used for speaker and speech feature extraction. X-vectors provide a fixed-length representation of the audio segment by capturing essential features through a DNN and pooling statistical summaries across frames. The process involves training a deep neural network to capture speaker or speech-related information from audio. Given an input audio signal $x[n]$, it is processed into overlapping frames represented by feature vectors: 
\[
X = \{x_1, x_2, \dots, x_T\}
\]
where $T$ is the total number of frames, and each $x_T$ is a $d$-dimensional feature vector.

Features of Each frame are passed through a DNN, transforming them into higher-level embeddings as follows:
\begin{equation}
h_i^{(l)} = f(W^{(l)} h_i^{(l-1)} + b^{(l)})
\end{equation}
where $W^{(l)}$ and $b^{(l)}$ are the weights and biases of the $l$-th layer, and $f$ is a nonlinear activation function.

The X-vector is then formed by concatenating the mean $\mu$ and standard deviation $\sigma$:
\begin{equation}
z = [\mu, \sigma]
\end{equation}

Finally, the X-vector is further refined through fully connected layers:
\begin{equation}
e = f(W_{\text{segment}} z + b_{\text{segment}})
\end{equation}


\begin{figure}[t]
  \centering
  \includegraphics[width=0.5\textwidth]{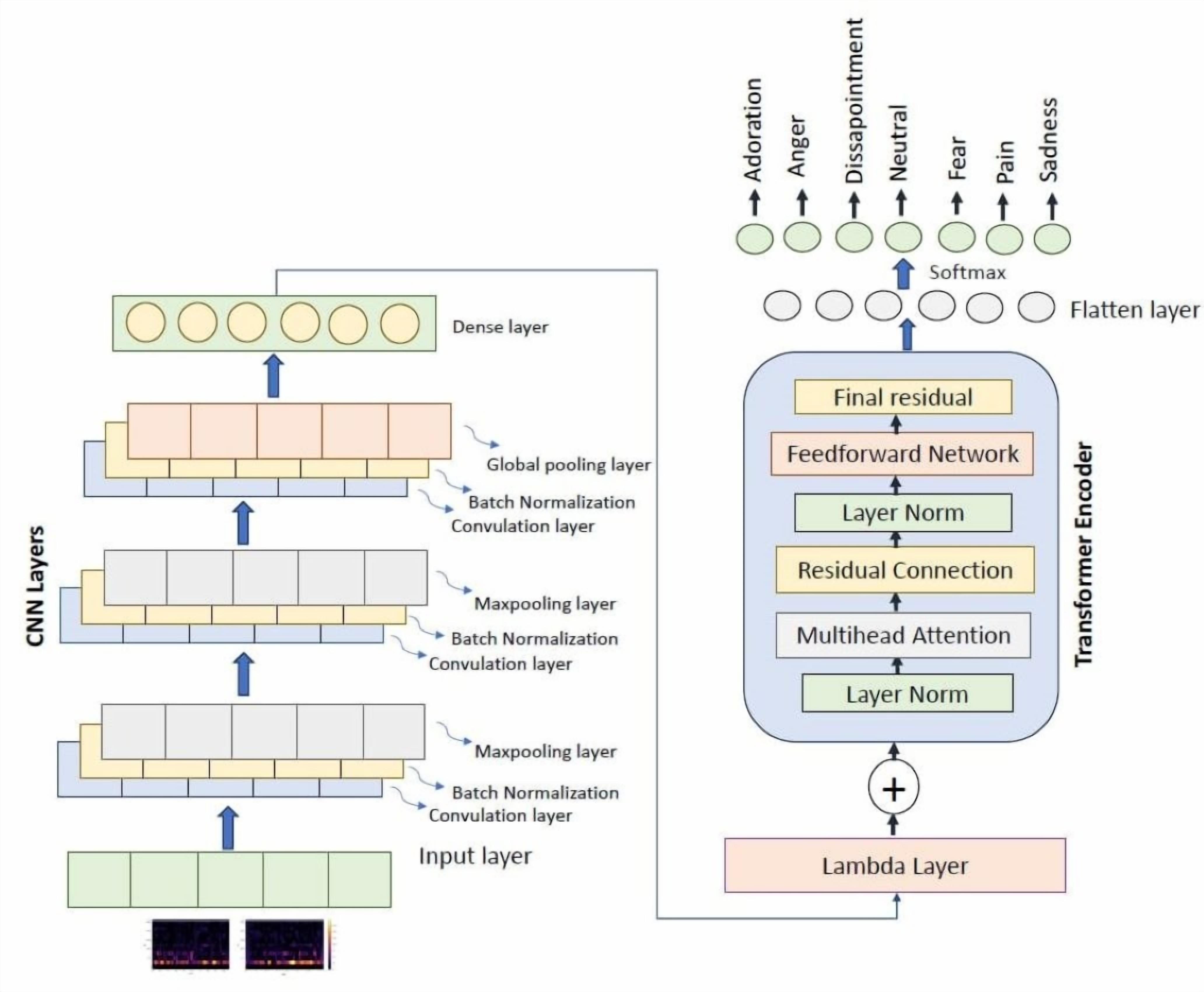}
  \caption{Architecture of the proposed EmoFormer network}
  \label{fig: Model Architecture}
   \vspace{-5pt}
\end{figure}


 
\subsection{Model Architecture}

The proposed EmoFormer network features a hybrid architecture that combines CNNs and transformer encoders to capture both local and global patterns in audio data for emotion recognition.

As shown in Fig.~\ref{fig: Model Architecture}, the model was tested with two types of input features: MFCC and X-vector. For MFCC, the input shape was (50,469), while for X-vectors it was (512,1). The architecture begins with four convolutional layers. The first CNN layer applies 16 filters with a 5x5 kernel and ReLU activation, followed by batch normalization and max-pooling to reduce spatial dimensions. The second and third CNN layers apply 32 and 64 filters, respectively, each using a 3x3 kernel, followed by batch normalization and max-pooling. The final CNN layer retains 64 filters and is followed by global average pooling. A dense layer with 64 units is then used to reduce dimensionality before passing the features to the transformer block.

\begin{table}[h]
\caption{Network architecture of the proposed EmoFormer model}
\begin{center}
\small
\begin{tabular}{|c|c|c|c|}
\hline
\textbf{Layer} & \textbf{Kernel Size} & \textbf{Input Shape} & \textbf{Output Shape} \\
\hline
Conv2D & (5, 5) & (13, 469, 1) & (13, 469, 16) \\
\hline
Conv2D & (3, 3) & (13, 469, 16) & (13, 469, 32) \\
\hline
Conv2D & (3, 3) & (13, 469, 32) & (6, 234, 32) \\
\hline
Conv2D & (3, 3) & (6, 234, 32) & (3, 117, 64) \\
\hline
Conv2D & (3, 3) & (3, 117, 64) & (1, 58, 64) \\
\hline
Conv2D & (3, 3) & (1, 58, 64) & (1, 58, 64) \\
\hline
Dense  & - & (1, 58, 64) & (64,) \\
\hline
Transformer Encoder & - & (64,) & (64,) \\
\hline
Flatten & - & (64,) & (64,) \\
\hline
Dense (Output) & - & (64,) & (7,) \\
\hline
\end{tabular}
\label{tab:cnn_transformer}
\end{center}
\end{table}

Following the CNN layers, a transformer encoder is applied. The encoder begins with layer normalization, followed by Multi-Head Attention with 8 heads to focus on different parts of the input sequence. Each attention head and the feed-forward layer have a dimensionality of 128. A dropout layer with a rate of 0.2 is used to reduce overfitting, followed by a residual connection and another layer normalization. Both normalization layers use an epsilon value of $1^{-6}$. Afterward, a dense layer with ReLU activation is followed by a dropout layer and another dense layer. The output is flattened into a 1D vector to be fed into the final dense layer, which applies a softmax activation function to produce predictions for the 5 to 23 emotion classes.
Table~\ref{tab:cnn_transformer} outlines the layer-wise architecture of the CNN-Transformer model.

\subsection{Experimental Setup}

In this work, various pre-processing techniques were applied to optimize data preparation for emotion recognition and improve model performance. Standardization was used to scale numerical features to a mean of 0 and a standard deviation of 1, ensuring that all features were on a similar scale and preventing features with larger magnitudes from dominating the learning process. The dataset was split into training and testing sets, with 70\% (2,621 samples) used for training and 30\% (1,124 samples) reserved for testing. Label encoding was applied to convert categorical variables into numerical values, assigning integer labels to each category. However, this approach may introduce an unintended ordinal relationship between categories.

The model was trained using the Adam optimizer and categorical cross-entropy as the loss function. For the MFCC-based model, early stopping was applied with patience of 10 epochs, and training was performed over 50 epochs with a batch size of 64. Similarly, the X-vector-based model used early stopping with patience of 5 epochs and was trained for 20 epochs with the same batch size. Both models monitored validation accuracy to control training and prevent overfitting, ensuring the models were effectively tuned to the data.





\section{Results}



This section presents the proposed network's performance across different sets of emotions, evaluated using precision, recall, F1-score, and accuracy. Additionally, the model's performance is compared to a baseline network in terms of accuracy percentage.

\begin{figure}[H]
\centering
\includegraphics[width=\columnwidth]{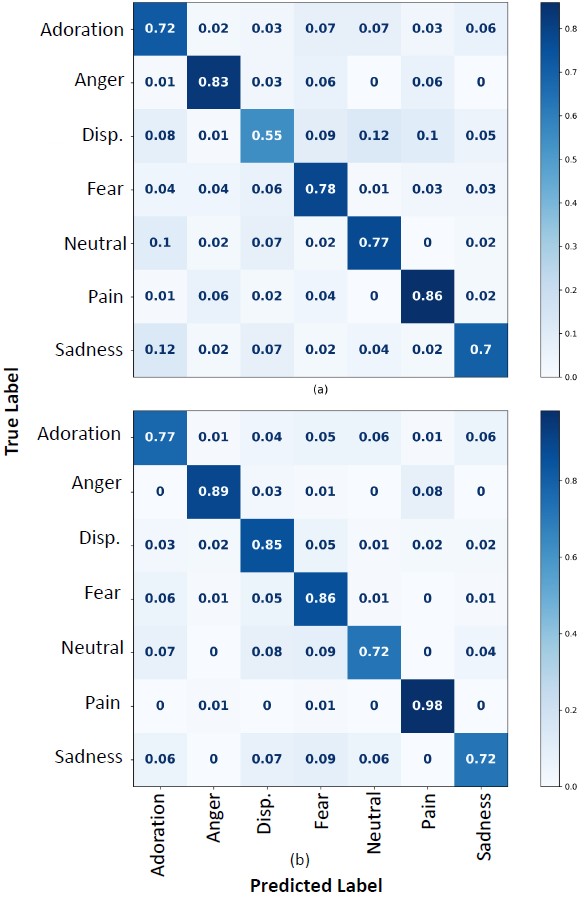}  
\caption{Confusion matrix generated using a) MFCC and b) X-vector}
\label{fig:mfcc_confusion}
\end{figure}

Figure~\ref{fig:mfcc_confusion} shows the confusion matrices for the proposed model when classifying seven emotions using MFCC and X-vector features. These matrices visually represent the model's classification performance, where diagonal values represent correct classifications and off-diagonal values indicate misclassifications. The MFCC-based model demonstrates strong performance in identifying emotions such as anger, fear, disappointment, and pain. The X-vector-based model also performs well in classifying anger and pain. However, the MFCC-based model consistently outperforms the X-vector model overall.

Table~\ref{tab1} provides a detailed comparison of the performance of the SER model in classifying seven emotions using MFCC, X-vector, and combined MFCC+X-vector features. The results are presented for both the models trained without data augmentation and those trained with augmentation.

For models trained without augmentation, the MFCC-based model significantly outperforms the X-vector-based model. The MFCC model achieves a precision of 0.45, recall of 0.41, and F1-score of 0.41. In contrast, the X-vector model shows lower performance with a precision of 0.22, recall of 0.22, and F1-score of 0.21.

When features are combined (MFCC+X-vector) without augmentation, the performance improves over X-vector alone, yielding a precision of 0.56, recall of 0.56, and F1-score of 0.55. However, it still lags behind the standalone MFCC model.

For models trained with augmentation, the MFCC model maintains its superior performance, with precision, recall, and F1-score values all reaching 0.83. The augmented X-vector model also improves the value of precision, recall, and F1-scores with 0.74 each. But it still falls behind the augmented MFCC model. The combination of MFCC+X-vector features with augmentation improves performance to a precision of 0.80, recall of 0.80, and an F1-score of 0.80.

Overall, as the MFCC features consistently outperformed the X-vector features, both with and without augmentation, MFCC features were used for the rest of the analysis in this study.

\begin{table}[h]
\caption{Comparison of SER Performances for classifying seven emotions using MFCC and X-vector features}
\begin{center}
\resizebox{\columnwidth}{!}{ 
\begin{tabular}{ccccc}
\hline\hline
\textbf{Feature} & \textbf{Augmentation} & \textbf{Precision} & \textbf{Recall} & \textbf{F1-score}\\
\hline\hline
MFCC & No  & 0.45 & 0.41 & 0.41 \\[0.5em] 
MFCC & Yes & 0.83 & 0.83 & 0.83 \\[0.5em] 
X-vector & No  & 0.22 &  0.22 & 0.21 \\[0.5em] 
X-vector & Yes & 0.74 & 0.74 & 0.74 \\[0.5em] 
MFCC+X-vector & No  & 0.56 & 0.56 & 0.55 \\[0.5em] 
MFCC+X-vector & Yes &  0.80 & 0.80 & 0.80\\[0.5em] 
\hline
\end{tabular}
}
\label{tab1}
\end{center}
\end{table}

The results obtained using MFCC features for different sets of emotions are presented in Table~\ref{tab2}. As seen in the table, when evaluating five emotions—adoration, anger, fear, neutral, and sadness—the model achieved an accuracy of 90\%, with a precision of 0.92, recall of 0.91, and an F1-score of 0.91, demonstrating strong performance.

\begin{table}[h]
\caption{Comparison of SER Performances in classifying different sets of emotions with augmented dataset}
\begin{center}
\resizebox{\columnwidth}{!}{ 
\begin{tabular}{ccccc}
\hline\hline
\textbf{No. of Emotions} & \textbf{Precision} & \textbf{Recall} & \textbf{F1-score} & \textbf{Accuracy (\%)}\\
\hline\hline \\[-1em]
5  & 0.92 & 0.91 & 0.91 & 90 \\[0.5em] 
7  & 0.83 & 0.83 & 0.83 & 83 \\[0.5em] 
10 & 0.75 & 0.72 & 0.72 & 72 \\[0.5em] 
23 & 0.66 & 0.65 & 0.64 & 65 \\[0.5em] 
\hline
\end{tabular}
}
\label{tab2}
\end{center}
\end{table}

%




Next, two additional emotions—disappointment and pain—were introduced to assess the model's performance. In this case, the accuracy dropped to 83\%, with an F1-score of 0.826, indicating that classification becomes more challenging as the number of emotions increases.

When the set was expanded to include guilt, disgust, and distress, the model's performance decreased further, with accuracy falling to 72\% and the F1-score to 0.719. Finally, when tasked with classifying 23 emotions, including a broader range such as amazement, confusion, serenity, and others, accuracy declined to 65\%, and the F1-score dropped to 0.64. This reduction in performance could be attributed to the overlapping nature of acoustic features among the emotions. Additionally, the computational time for 7 emotions is around 1.34s, and we also calculated the model's Multiply Accumulate(MAC) value, which is 35041444. In summary, the results indicate that the model performs significantly better with a smaller set of emotions. As the number of emotions increases, the classification task becomes more complex, leading to lower accuracy and F1 scores.

\begin{table}
\caption{Comparison with baseline networks and ablation study for the proposed network}
\begin{center}
\large
\resizebox{\columnwidth}{!}{ 
\begin{tabular}{ccccc}
\hline\hline
\textbf{Model} & \textbf{Precision} & \textbf{Recall} & \textbf{F1-score} & \textbf{Accuracy(\%)}\\
\hline\hline
\\[-1em] 
CNN-Model\cite{chauhan2021speech} & 0.69 & 0.70 & 0.71 & 70 \\[0.5em]
Transformer-LSTM \cite{andayani2022hybrid} & 0.49 & 0.39 & 0.38 & 39 \\[0.5em]
Stacked-LSTM & 0.64 & 0.60 & 0.61 & 60 \\[0.5em]
Stacked-GRU & 0.65 & 0.63 & 0.64 & 63 \\[0.5em]
Transformer+2-Bi-LSTM & 0.69 & 0.68 & 0.68 & 68 \\[0.5em]
\textbf{Proposed EmoFormer} & \textbf{0.83} & \textbf{0.83} & \textbf{0.83} & \textbf{83} \\[0.5em]
\hline
\end{tabular}
}
\label{tab3}
\end{center}
\end{table}

This study compares the proposed network's SER performance against two baseline networks: one based on CNN \cite{chauhan2021speech} and another using a hybrid Transformer-LSTM architecture \cite{andayani2022hybrid}. The performance evaluation, conducted on seven emotions, is presented in Table~\ref{tab3}. The CNN model \cite{chauhan2021speech} achieved an accuracy of 70\%, while the hybrid Transformer-LSTM model \cite{andayani2022hybrid} performed significantly worse, with only 39\% accuracy.

In contrast, the proposed network showed promising results. When using Stacked GRU and Stacked LSTM layers, the model achieved 63\% and 60\% accuracy, respectively. Performance improved with a Transformer network incorporating Bi-LSTM layers, reaching 68\% accuracy. The most significant improvement occurred when Bi-LSTM layers were replaced with CNN in the Transformer-BiLSTM model, resulting in an accuracy of 83\%. Additionally, the EmoFormer network attained strong precision (0.8293), recall (0.8265), and F1-score (0.8255). These results not only highlight the superior performance of the proposed model compared to the baseline architectures tested in this study but also show that it substantially outperforms models from previous research.


\section{Conclusion}
This study introduces a hybrid model, known as EmoFormer, for SER, combining CNN and Transformer encoders. While prior studies have relied on datasets that were not entirely text-independent, our approach uniquely utilizes a dataset that is fully text-independent. The model was evaluated using MFCC and x-vector features on an augmented speech dataset. The results demonstrate that the MFCC feature-based model consistently outperformed the x-vector model across various emotion sets. Notably, the model achieved a peak accuracy of 90\% when classifying five emotions, although performance gradually declined as the number of emotions increased, reaching 65\% for 23 emotions due to the complexity and overlap of acoustic features. Compared to baseline networks, including CNN and Transformer-LSTM architectures, the proposed model showed superior performance. It achieved 83\% accuracy for seven emotions, along with high precision, recall, and F1-scores, demonstrating its effectiveness in emotion classification. Future work could focus on improving the classification of subtle emotions by incorporating multimodal data, such as integrating speech with visual cues like facial expressions or body language. Additionally, exploring more advanced Transformer architectures or ensemble models may help address performance declines as the number of emotions increases.


\bibliographystyle{IEEEtran}
\bibliography{Manuscript.bib}

\def\authornoop#1{}
\begin{thebibliography}{10}
\providecommand{\url}[1]{#1}
\csname url@samestyle\endcsname
\providecommand{\newblock}{\relax}
\providecommand{\bibinfo}[2]{#2}
\providecommand{\BIBentrySTDinterwordspacing}{\spaceskip=0pt\relax}
\providecommand{\BIBentryALTinterwordstretchfactor}{4}
\providecommand{\BIBentryALTinterwordspacing}{\spaceskip=\fontdimen2\font plus
\BIBentryALTinterwordstretchfactor\fontdimen3\font minus \fontdimen4\font\relax}
\providecommand{\BIBforeignlanguage}[2]{{%
\expandafter\ifx\csname l@#1\endcsname\relax
\typeout{** WARNING: IEEEtran.bst: No hyphenation pattern has been}%
\typeout{** loaded for the language `#1'. Using the pattern for}%
\typeout{** the default language instead.}%
\else
\language=\csname l@#1\endcsname
\fi
#2}}
\providecommand{\BIBdecl}{\relax}
\BIBdecl

\bibitem{ramakrishnan2013speech}
S.~Ramakrishnan and I.~M. El~Emary, ``Speech emotion recognition approaches in human computer interaction,'' \emph{Telecommunication Systems}, vol.~52, pp. 1467--1478, 2013.

\bibitem{avro2024Emotech}
S.~Bin Habib~Avro, T.~Taher, and N.~Mamun, ``Emotech: A multi-modal speech emotion recognition using multi-source low-level information with hybrid recurrent network,'' in \emph{International Conference on Signal Processing, Information, Communication and Systems}.\hskip 1em plus 0.5em minus 0.4em\relax IEEE, 2024, pp. 1--5.

\bibitem{wani2021comprehensive}
T.~M. Wani, T.~S. Gunawan, S.~A.~A. Qadri, M.~Kartiwi, and E.~Ambikairajah, ``A comprehensive review of speech emotion recognition systems,'' \emph{IEEE access}, vol.~9, pp. 47\,795--47\,814, 2021.

\bibitem{koduru2020feature}
A.~Koduru, H.~B. Valiveti, and A.~K. Budati, ``Feature extraction algorithms to improve the speech emotion recognition rate,'' \emph{International Journal of Speech Technology}, vol.~23, no.~1, pp. 45--55, 2020.

\bibitem{mao2019revisiting}
S.~Mao, D.~Tao, G.~Zhang, P.~Ching, and T.~Lee, ``Revisiting hidden markov models for speech emotion recognition,'' in \emph{ICASSP 2019-2019 IEEE International Conference on Acoustics, Speech and Signal Processing (ICASSP)}.\hskip 1em plus 0.5em minus 0.4em\relax IEEE, 2019, pp. 6715--6719.

\bibitem{kerkeni2019automatic}
L.~Kerkeni, Y.~Serrestou, M.~Mbarki, K.~Raoof, M.~A. Mahjoub, and C.~Cleder, ``Automatic speech emotion recognition using machine learning,'' \emph{Social Media and Machine Learning}, 2019.

\bibitem{mirsamadi2017automatic}
S.~Mirsamadi, E.~Barsoum, and C.~Zhang, ``Automatic speech emotion recognition using recurrent neural networks with local attention,'' in \emph{2017 IEEE International conference on acoustics, speech and signal processing (ICASSP)}.\hskip 1em plus 0.5em minus 0.4em\relax IEEE, 2017, pp. 2227--2231.

\bibitem{zhao2019speech}
J.~Zhao, X.~Mao, and L.~Chen, ``Speech emotion recognition using deep 1d \& 2d cnn lstm networks,'' \emph{Biomedical signal processing and control}, vol.~47, pp. 312--323, 2019.

\bibitem{anvarjon2020deep}
T.~Anvarjon, Mustaqeem, and S.~Kwon, ``Deep-net: A lightweight cnn-based speech emotion recognition system using deep frequency features,'' \emph{Sensors}, vol.~20, no.~18, p. 5212, 2020.

\bibitem{sarker2023text}
S.~Sarker, K.~Akter, and N.~Mamun, ``A text independent speech emotion recognition based on convolutional neural network,'' in \emph{2023 International Conference on Electrical, Computer and Communication Engineering (ECCE)}.\hskip 1em plus 0.5em minus 0.4em\relax IEEE, 2023, pp. 1--4.

\bibitem{he2023multiple}
Y.~He, N.~Minematsu, and D.~Saito, ``Multiple acoustic features speech emotion recognition using cross-attention transformer,'' in \emph{ICASSP 2023-2023 IEEE International Conference on Acoustics, Speech and Signal Processing (ICASSP)}.\hskip 1em plus 0.5em minus 0.4em\relax IEEE, 2023, pp. 1--5.

\bibitem{morais2022speech}
E.~Morais, R.~Hoory, W.~Zhu, I.~Gat, M.~Damasceno, and H.~Aronowitz, ``Speech emotion recognition using self-supervised features,'' in \emph{ICASSP 2022-2022 IEEE International Conference on Acoustics, Speech and Signal Processing (ICASSP)}.\hskip 1em plus 0.5em minus 0.4em\relax IEEE, 2022, pp. 6922--6926.

\bibitem{aftab2022light}
A.~Aftab, A.~Morsali, S.~Ghaemmaghami, and B.~Champagne, ``Light-sernet: A lightweight fully convolutional neural network for speech emotion recognition,'' in \emph{ICASSP 2022-2022 IEEE international conference on acoustics, speech and signal processing (ICASSP)}.\hskip 1em plus 0.5em minus 0.4em\relax IEEE, 2022, pp. 6912--6916.

\bibitem{chauhan2021speech}
K.~Chauhan, K.~K. Sharma, and T.~Varma, ``Speech emotion recognition using convolution neural networks,'' in \emph{2021 international conference on artificial intelligence and smart systems (ICAIS)}.\hskip 1em plus 0.5em minus 0.4em\relax IEEE, 2021, pp. 1176--1181.

\bibitem{andayani2022hybrid}
F.~Andayani, L.~B. Theng, M.~T. Tsun, and C.~Chua, ``Hybrid lstm-transformer model for emotion recognition from speech audio files,'' \emph{IEEE Access}, vol.~10, pp. 36\,018--36\,027, 2022.

\end{thebibliography}

\end{document}